\documentclass[preprint,aps,floats,nofootinbib]{revtex4}

\input epsf.tex
\def\DESepsf(#1 width #2){\epsfxsize=#2 \epsfbox{#1}}

\usepackage{graphicx}

\newcommand{\be}{\begin{equation}}
\newcommand{\ee}{\end{equation}}
\newcommand{\bea}{\begin{eqnarray}}
\newcommand{\eea}{\end{eqnarray}}

\def\thebibliography#1{\centerline{\bf REFERENCES}
  \list{[\arabic{enumi}]}{\settowidth\labelwidth{[#1]}\leftmargin
  \labelwidth\advance\leftmargin\labelsep\usecounter{enumi}}
\def\newblock{\hskip .11em plus .33em minus -.07em}\sloppy
  \clubpenalty4000\widowpenalty4000\sfcode`\.=1000\relax}

\begin{document}
\preprint{\vbox{\hbox{}\hbox{}\hbox{}}} 

\vspace*{0.5cm}

\title{$B_s - \bar B_s$ Mixing and Its Implication for $b \to s$ Transitions \\
in Supersymmetry}

\author{ \vspace{0.5cm}
Richard Arnowitt$^1$\footnote{arnowitt@physics.tamu.edu},
~Bhaskar~Dutta$^1$\footnote{dutta@physics.tamu.edu}, ~Bo
Hu$^2$\footnote{bohu@ncu.edu.cn} ~ and ~
Sechul~Oh$^3$\footnote{scoh@phya.yonsei.ac.kr}}

\affiliation{ \vspace{0.3cm}
$^1$Department of Physics, Texas A\&M University, College Station, TX, 77845, USA \\
$^2$Department of Physics, Nanchang University, Jiangxi 330047, China \\
$^3$Natural Science Research Institute, Yonsei University, Seoul
120-479, Korea} \vspace*{0.5cm}

\begin{abstract}
\noindent
We investigate the effect of the current measurement of the neutral $B_s$ meson
mass difference, $\Delta M_{B_s}$, on  SUGRA models which have non-zero values
of the soft breaking terms $(m^2_{LL,RR})_{23}$ and $A^{u,d}_{23}$ at the GUT
scale. We use non-zero values of these parameter to explain the $B\to K\pi$ puzzle
and find that even after satisfying the experimental result on $\Delta M_{B_s}$
and the branching ratio (BR) of $b\to s\gamma$ we still can explain the puzzle.
Further we show that in this parameter space it is possible to accommodate the
large BR of $B\to \eta' K$ and the current experimental data for CP asymmetries
of $B\to \eta' K^0$ and $B\to \phi K^0$.
The predicted value of $\sin (2\beta^{\rm eff})_{\eta' K^0}$ is  about $0.52-0.67$.
\end{abstract}
\maketitle

Flavor changing $b \to s$ transitions are particularly interesting for new physics (NP) searches
using $B$ meson decays.  In the standard model (SM) these transitions can occur only at the
loop-level so that they are particulary sensitive to NP effects.  So far, a few {\it possible}
indications to NP effects through $b \to s$ transitions have been reported by experimental
collaborations such as BaBar and Belle.  Among them is the recent $B \to K\pi$ puzzle: i.e.,
discrepancies between the SM predictions and the experimental results for the direct and
mixing-induced CP asymmetries and the branching ratios (BRs) in $B \to K\pi$ modes whose dominant
quark level processes are $b \to s q \bar q$ $(q = u,d)$~\cite{HFAG,Kim:2005jp,Arnowitt:2005qz}.
The measurements of the CP asymmetries in $B_d \to \eta' K$ and $B_d \to \phi K$ modes as well as
the rather large BR for $B \to \eta' K$ and $B \to \eta K$ also have drawn a lot of attention,
due to their possible deviation from the SM predictions ~\cite{HFAG,Dutta:2003ha}.  The (dominant)
subprocess of these modes is the $b \to s s \bar s$ transition.

Recently, the CDF collaboration has reported a new result for another interesting observable
relevant to the $b \to s$ transition: the mass difference between the neutral $B_s$ states that
characterizes the $B_s - \bar B_s$ mixing phenomenon.
The CDF result is~\cite{CDF}
\begin{equation}
\Delta M_{B_s} = 17.33^{+0.42}_{-0.21} ({\rm stat.}) \pm 0.07 ({\rm syst.}) ~~{\rm ps}^{-1}~.
\label{CDF}
\end{equation}
The D$\emptyset$ collaboration has also recently provided a new result~\cite{Abazov:2006dm}:
\begin{equation}
17 ~~{\rm ps}^{-1} < \Delta M_{B_s} < 21 ~~{\rm ps}^{-1} ~~ (90\% ~{\rm C.L.}) ~.
\end{equation}
These experimental results are consistent with the SM estimation.
Therefore, these new experimental results are expected to provide important constraints
on  NP  beyond the SM~\cite{Deshpande:1996yt}.
Motivated by these new results, some theoretical studies have been done to search for
NP effects~\cite{Ciuchini:2006dx,Endo:2006dm,Blanke:2006ig,Ligeti:2006pm,Foster:2006ze,
Cheung:2006tm,Ball:2006xx,Khalil:2006pv,Grossman:2006ce,Baek:2006fq,He:2006bk}.

In the SM, the mass difference in the $B_s$ system is given by
\begin{equation}
\Delta M^{\rm SM}_{B_s} = \frac{G_F^2 M_W^2}{6 \pi^2} M_{B_s} \hat{\eta}_B \hat{B}_{B_s}
  f_{B_s}^2 \left| V_{tb} V_{ts}^* \right|^2 S_0 (x_t) ~,
\label{DeltaMsSM}
\end{equation}
where the NLO short-distance QCD correction gives $\hat{\eta}_B = 0.552$ and $S_0 (x_t) =
2.463$~\cite{Buchalla:1995vs}.  The non-perturbative quantities $\hat{B}_{B_s}$ and $f_{B_s}$ are
the bag parameter and the decay constant, respectively. The best fit for $\Delta M^{\rm SM}_{B_s}$
is given by~\cite{Bona:2005eu,Charles:2004jd}
\begin{eqnarray}
\Delta M^{\rm SM}_{B_s} = 21.5 \pm 2.6 ~~ {\rm ps}^{-1} ~~~ [{\rm
UTfit}]~, ~~~~ \Delta M^{\rm SM}_{B_s} = 21.7^{+5.9}_{-4.2} ~~ {\rm
ps}^{-1} ~~~ [{\rm CKMfitter}]~, \label{DeltaMsFIT}
\end{eqnarray}
In a recent paper~\cite{Ball:2006xx}, this mass difference is found to be $23.4 \pm 3.8~~{\rm ps}^{-1}$
using HPQCD and JLQCD data for $f_{B_s}\sqrt{\hat{B}_{B_s}}$.

In this letter, we  study the neutral $B_s$ meson mixing  effect in supersymmetry (SUSY): specifically
in the supergravity (SUGRA) model.  Then, using the constraints obtained from $\Delta M_{B_s}$, we
focus on how to resolve all the {\it possible} current anomalies observed in hadronic $B \to PP$ ($P$
denotes a pseudoscalar meson) decays through the $b \to s$ transitions, such as $B \to K \pi$,
$B \to \eta' K$. The current experimental data is listed in Table 1.

We consider the SUGRA model with the simplest possible non-universal soft terms which is the simplest
extension of the minimal SUGRA (mSUGRA) model.  In the SUGRA model, the superpotential and soft SUSY
breaking terms at the grand unified theory (GUT) scale are given by
\begin{eqnarray}
{\cal W} &=& Y^U Q H_2 U + Y^D Q H_1 D + Y^L L H_1 E + \mu H_1 H_2 ,
\nonumber \\
{\cal L}_{\rm soft} &=& - \sum_i m_i^2 |\phi_i|^2
 - \left[ {1 \over 2} \sum_{\alpha} M_{\alpha} \bar \lambda_{\alpha}
 \lambda_{\alpha} + B \mu H_1 H_2 \right.  \nonumber \\
&\mbox{}& \left. + (A^U Q H_2 U + A^D Q H_1 D + A^L L H_1 E)
  + {\rm H.c.} \right],
\end{eqnarray}
where $E$, $U$ and $D$ are respectively the lepton, up-quark and down-quark singlet superfields, $L$
and $Q$ are the SU$(2)_L$ doublet lepton and quark superfields, and $H_{1,2}$ are the Higgs doublets.
$\phi_i$ and $\lambda_{\alpha}$ denote all the scalar fields and gaugino fields, respectively.

The SUSY contributions appear at loop order. In our calculation, we do not use the mass insertion
approximation, but rather do a complete calculation~\cite{Bertolini:1990if,Gabbiani:1996hi}.
We assume the breakdown of the universality to accommodate the $b \to s$ transition data.  While we
satisfy this data, we also have to be careful to satisfy other data, e.g., $b\rightarrow s\gamma$.

We use the following boundary conditions at the GUT scale:
\begin{equation}
\left( m^{2}_{(Q_{LL},U_{RR},D_{RR})} \right)_{ij}
 = m_0^2 \left[ \delta_{ij} + \left( \Delta_{(Q_{LL},U_{RR},D_{RR})} \right)_{ij} \right] ~,
 ~~~ A_{ij}^{(u,d)} = A_0 \left( Y_{ij}^{(u,d)} +\Delta A_{ij}^{(u,d)} \right) ~,
\end{equation}
where $i,j=1,2,3$ are the generation indices.  The SUSY parameters can have phases at the GUT scale:
$M_{i}=|M_{1/2}| e^{i\theta_i}$ (the gaugino masses for the $U(1)$, $SU(2)$ and $SU(3)$ groups,
$i=1,2,3$), $A_0=|A_0| e^{i\alpha_{_A}}$ and $\mu=|\mu| e^{i\theta_\mu}$.  However, we can set one
of the gaugino phases to zero and we choose $\theta_2 =0$. The electric dipole moments (EDMs) of the
electron and neutron can now allow the existence of large phases in the theory~\cite{AAB1}.  In our
calculation, we use $O(1)$ phases but calculate the EDMs to make sure that current bounds
($|d_e| <1.2 \times 10^{-27}$ecm~\cite{Regan:2002ta} and
$|d_n| <6.3 \times 10^{-26}$ecm~\cite{Harris:1999jx}) are satisfied.

We evaluate the squark masses and mixings at the weak scale by using the above boundary conditions
at the GUT scale.  The RGE evolution mixes the non-universality of type LR (A terms) via
$d {m_Q}^2_{LL,RR} /dt \propto A_{u(d)}^{\dagger}A_{u(d)}$ terms and creates new LL and RR
contributions at the weak scale.  We then evaluate the Wilson coefficients from all these new
contributions.  We have both chargino and gluino contributions arising due to the LL, LR, RR up type
and down type squark mixing.  These contributions affect the following Wilson coefficients $C3-C9$,
$C_{7\gamma}$ and $C_{8g}$.  The chargino contributions affect mostly the electroweak penguins
(C7 and C9) and the dipole penguins, where as the gluino penguin has the largest contribution to the
dipole penguins due to the presence of an enhancement factor $m_{\tilde g}/m_b$ (The gluino
contribution also affects the QCD penguins).  We include all contributions in our calculation.

For calculation of the relevant hadronic matrix elements, we adopt the QCD improved factorization.
This approach allows us to include the possible non-factorizable contributions, such as vertex
corrections, penguin corrections, hard spectator scattering contributions, and weak annihilation
contributions~\cite{bbns}.

\begin{table}
\caption{Experimental data on the CP-averaged branching ratios
($\bar {\cal B}$ in units of $10^{-6}$), the direct CP asymmetries
(${\cal A}_{CP}$), and the effective $\sin (2\beta)$ ($\beta$ is the
angle of the unitarity triangle) for $B \to PP$ decays~\cite{HFAG}.}
\smallskip
\begin{tabular}{|c|c||c|c|}
\hline
 BR & Average & CP asymmetry & Average  \\
\hline $\bar {\cal B}(B^{\pm} \to K^0 \pi^{\pm})$ & $24.1 \pm 1.3$
 & ${\cal A}_{CP}(K^0 \pi^{\pm})$ & $-0.02 \pm 0.04$  \\
$\bar {\cal B}(B^{\pm} \to K^{\pm} \pi^0)$ & $12.1 \pm 0.8$
 & ${\cal A}_{CP}(K^{\pm} \pi^0)$ & $+0.04 \pm 0.04$  \\
$\bar {\cal B}(B^0 \to K^{\pm} \pi^{\mp})$ & $18.9 \pm 0.7$
 & ${\cal A}_{CP}(K^{\pm} \pi^{\mp})$ & $-0.115 \pm 0.018$  \\
$\bar {\cal B}(B^0 \to K^0 \pi^0)$ & $11.5 \pm 1.0$
 & ${\cal A}_{CP}(K^0 \pi^0)$ & $+0.001 \pm 0.155$  \\
 &  & $\sin (2\beta^{\rm eff})_{K_s \pi^0}$ & $+0.34 \pm 0.29$  \\
\hline $\bar {\cal B}(B^{\pm} \to \eta' K^{\pm})$ &
$69.7^{+2.8}_{-2.7}$
 & $\sin (2\beta^{\rm eff})_{\eta' K^0}$ & $+0.50 \pm 0.09$  \\
 &  & $\sin (2\beta^{\rm eff})_{\phi K^0}$ & $+0.47 \pm 0.19$  \\
\hline
\end{tabular}
\label{table:1}
\end{table}

The neutral $B$ meson mass difference involves gluino and chargino
diagrams in SUSY~\cite{Gerard:1984vb}. In mSUGRA, with universal
boundary condition, the chargino diagram has the dominant
contribution. Once we introduce mixing in the (2,3)-sector of the
$m^2_{LL,RR}$ or $A_{LR}$ soft breaking terms, the mass difference
gets enhanced and we  get large contributions from the gluino
diagrams. The $B\to\pi K$ puzzle can not be solved using just the
mSUGRA boundary condition. In order to explain the $B\to K\pi$
puzzle, we have noticed that the flavor violating terms in the
(2,3)-sector of the soft breaking masses are
needed~\cite{Arnowitt:2005qz}.

In order to investigate the effect of the neutral $B_s$ mixing on the $b \to s$ transitions, we first
try to fit the $B\to K\pi$ data using $A_{23}^{u,d}$, $(m^2_{LL,RR})_{23}$ at the GUT scale.  The
constraint from the BR of $b\to s\gamma$ is also included.  We vary $m_{1/2}$ in the range $(350-500)$
GeV [corresponding to gluino mass of $(1-1.5)$ TeV], $A_0=-800$ GeV, $\Delta_{(Q_{LL},U_{RR},D_{RR})}
\sim 0-0.3$, $\Delta A_{23}^{(u,d)} =0-0.3$, $m_0=300$ GeV and $\tan\beta$=40.  The $\Delta$'s also
have O(1) phases.  The magnitudes of $\Delta$'s get reduced at the weak scale compared to the GUT scale
since the squark masses get a contribution from $m_{1/2}$ in the RGEs.

In Fig 1, we plot  ${\cal A}_{CP} (K^{\pm} \pi^{\mp})$ versus $\Delta M_{B_s}/\Delta M_{B_d}$, where
$\Delta M_{B_d}$ is the mass difference between the neutral $B_d$ states.  The experimental value for
the $\Delta M_{B_d}$ is $0.507\pm 0.005$ ps$^{-1}$.
In the SM,
\begin{equation}
{{\Delta M_s^{\rm SM}}\over{\Delta M_d^{\rm SM}}}
={M_{B_s}\over M_{B_d}}\xi^2 \left| {V_{ts}\over V_{td}} \right|^2 ~,
\end{equation}
where $\xi \equiv{{f_{B_s}\sqrt{\hat{B}_s}}\over
f_{B_d}\sqrt{\hat{B}_d}}$. In the plot, we used $\xi =1.18$ and the
CKM phase $\gamma=61.1^{\circ}\pm 4.5^{\circ}$~\cite{Bona:2005eu}.
We find that the 2$\sigma$ experimental range about the central
value of the ratio $\Delta M_{B_s}/\Delta M_{B_d} =34.66$ rules out
a lot of model points.  In order to extract the valid points, we
include the error of $\xi =1.23 \pm 0.06$~\cite{Blanke:2006ig}
(consistent with the value of $\xi = 1.21^{+0.047}_{-0.035}$ using
the JLQCD and the HPQCD calculations in Ref.~\cite{Ball:2006xx}),
and calculate the BRs and the CP asymmetries of different $B \to
K\pi$ modes.  We also calculate the BR of $B\to \eta^{'}K$ and $\sin
(2\beta^{\rm eff})_{\eta' K^0}$ as well as $\sin (2\beta^{\rm
eff})_{\phi K^0}$ for the allowed model points.

\begin{figure}
    \centerline{ \DESepsf(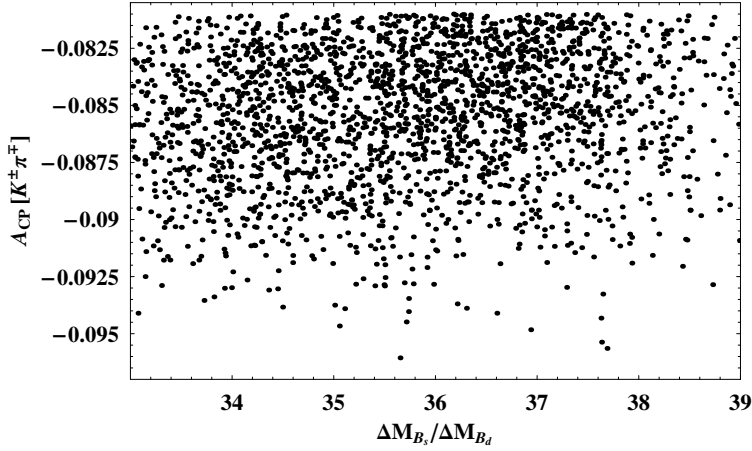 width 10cm)}
    \vspace{0.1cm}
    \caption{$A_{CP}(K^{\pm} \pi^{\mp})$ versus $\Delta M_{B_s}/\Delta M_{B_d}$
     in the SUGRA model. The parameters are described in the text.}
\label{fig:fig1}
\end{figure}

\begin{figure}
    \centerline{ \DESepsf(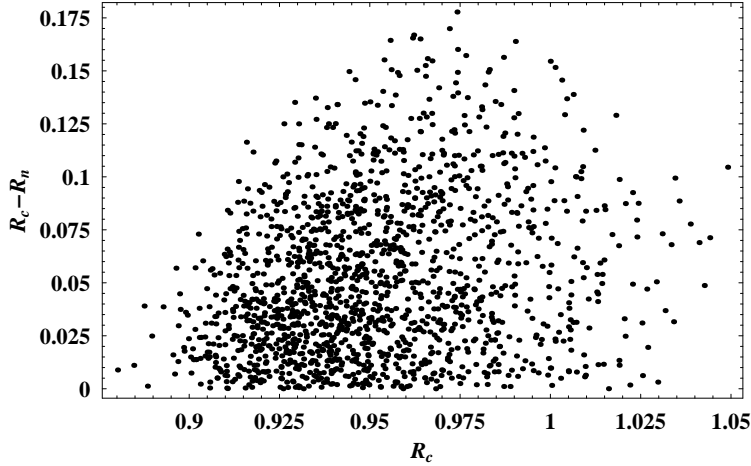 width 10cm)}
    \vspace{0.1cm}
    \caption{$R_c-R_n$ versus $R_c$ in the SUGRA model.}
\label{fig:fig2}
\end{figure}

The recent experimental data for the CP-averaged BRs of $B \to K\pi$ may indicate a possible deviation
from the prediction of the SM:
\begin{eqnarray}
R_c \equiv \frac{2\bar {\cal B}(B^{\pm} \to K^{\pm} \pi^0)}
 {\bar {\cal B}(B^{\pm} \to K^0 \pi^{\pm})}
 = 1.00 \pm 0.09 ~, ~~~~
R_n \equiv \frac{\bar {\cal B}(B^0 \to K^{\pm} \pi^{\mp})}
 {2\bar {\cal B}(B^0 \to K^0 \pi^0)}
 = 0.79 \pm 0.08 ~.
\label{RcRn}
\end{eqnarray}
It has been claimed that within the SM, $R_c \approx R_n$~\cite{Buras:2004th,Buras:2003dj}.  But, the
data show the pattern $R_c > R_n$, which would indicate the enhancement of the electroweak (EW) penguin
and/or the color-suppressed tree contributions~\cite{Kim:2005jp}.  In Fig. 2, we plot $R_c-R_n$ versus
$R_c$ and find that $R_c > R_n$ can be satisfied.

\begin{figure}
    \centerline{ \DESepsf(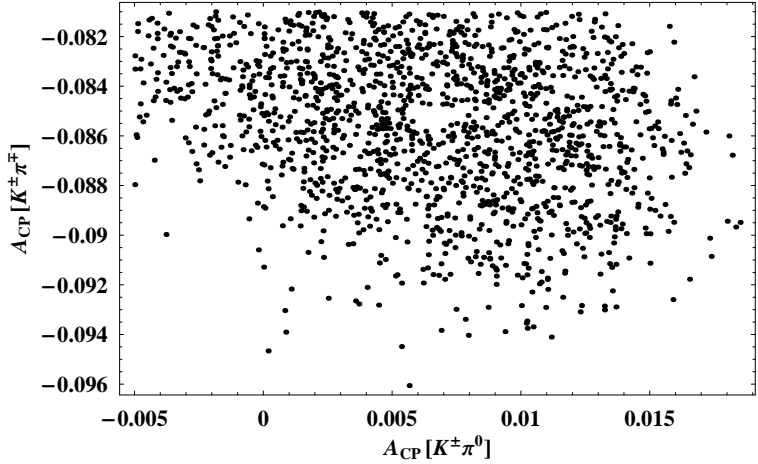 width 10cm)}
    \vspace{0.1cm}
    \caption{$A_{CP}(K^{\pm} \pi^0)$ versus $A_{CP}(K^{\pm} \pi^{\mp})$
     in the SUGRA model.}
\label{fig:fig3}
\end{figure}

\begin{figure}
    \centerline{ \DESepsf(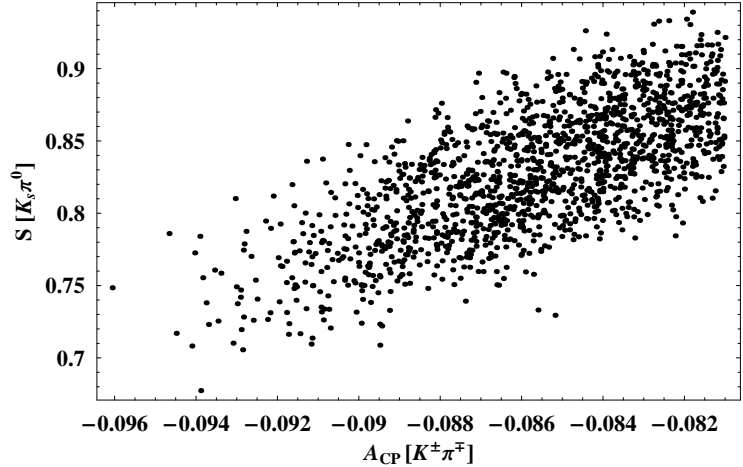 width 10cm)}
    \vspace{0.1cm}
    \caption{$\sin (2\beta^{\rm eff})_{K_s \pi^0}$ versus
     $A_{CP}(K^{\pm} \pi^{\mp})$ in the SUGRA model.}
\label{fig:fig4}
\end{figure}

Also, in the conventional prediction of the SM, ${\cal A}_{CP}(K^{\pm} \pi^0)$ is expected to be
almost the same as ${\cal A}_{CP}(K^{\pm} \pi^{\mp})$.  In particular, they would have the {\it same}
sign.  However, the current data show that ${\cal A}_{CP}(K^{\pm} \pi^0)$ differs by 3.5$\sigma$ from
${\cal A}_{CP}(K^{\pm} \pi^{\mp})$.  In Fig. 3, we plot ${\cal A}_{CP}(K^{\pm} \pi^0)$ versus
${\cal A}_{CP}(K^{\pm} \pi^{\mp})$ and find that the signs can be different for the points allowed by
the neutral B mixing data.

The predicted $\sin (2\beta^{\rm eff})_{K_{_S} \pi^0}$ is shown in Fig. 4.  We find that the minimum
value is 0.7.  The present experimental data still have large errors so that future results will
confirm/rule out our model.

\begin{figure}
    \centerline{ \DESepsf(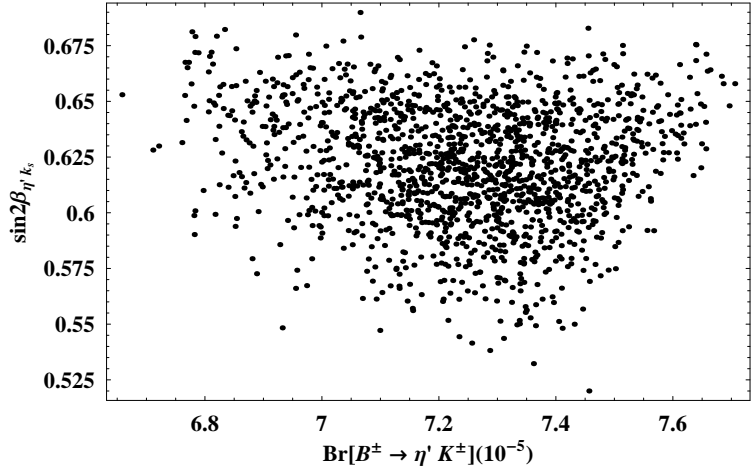 width 10cm)}
    \vspace{0.1cm}
    \caption{$\sin (2\beta^{\rm eff})_{\eta' K^0}$ versus
     the BR of $B^{\pm} \to \eta' K^{\pm}$
     in the SUGRA model.}
\label{fig:fig5}
\end{figure}

The experimental BRs of ${\cal B}(B\rightarrow\eta' K)$ are large
compared to the conventional SM predictions.  In Fig. 5, we plot
$\sin (2\beta^{\rm eff})_{\eta' K^0}$ versus the BR of $B^{\pm}
\to\eta' K^{\pm}$.  These decay modes get SUSY contributions since
we are using non-zero values of $(m^2_{LL,RR})_{23}$ and $A_{23}$
and the BR gets enhanced.
The values of $\sin(2\beta^{\rm eff})_{\eta'K^0}$ is  allowed by the
experimental value which has a smaller error than that of
$\sin (2\beta^{\rm eff})_{\phi K^0}$.
The value of $\sin (2\beta^{\rm eff})_{\phi K^0}$ is
predicted to be around $(0.55-0.70)$ and the BRs of $B^{\pm}\to
\phi K^{\pm}$ and $B^0\to \phi K^0$ are around $(7-9) \times
10^{-6}$ and $(6.5-8.5) \times 10^{-6}$ respectively in our
calculation and ${\cal B}(b\to s\gamma)$ is $(2-4.5) \times
10^{-4}$.  We also find that ${\cal B}(B\rightarrow\eta K)$ is
around $3\times 10^{-6}$. The CP asymmetries for $B^{\pm}\to\phi K^{\pm}$
and $B^{\pm}\to\eta^{(')} K^{\pm}$
are $-0.1$ to 0.1 and close to 0, respectively.
The Arg[$M(12)_{B_s}$] is less than $5^{\circ}$ for our model points.

It is possible to obtain a fit for the experimental results even
without using $m_{LL}^2$ contribution at al.  The nonzero values of
$A^{u,d}_{23}$ parameters generate the dipole penguin and the
($Z$-mediated) electroweak penguin diagrams.  As a representative
example, we present the BRs and CP asymmetries for a specific model
point in Table II to show that all these different experimental
results can be explained by one model point using just
$A^{u,d}_{23}$.  The parameters of the model point are given at the
GUT scale by $m_{1/2} =450$ GeV, $m_0 =350$ GeV, $A_0 = -800$ GeV,
$\Delta A^d_{23}=0.1 ~e^{-2.0 i}$, $\Delta A^u_{23}=0.48 ~e^{1.1
i}$, and we choose $\tan\beta =40$. We find that the BRs and CP
asymmetries are all within one sigma of the experimental results
except for $\sin (2\beta^{\rm eff})_{K_s \pi^0}$ which is about
1.6$\sigma$ away (this deviation is lowered when we include
$m^2_{LL}$ contribution). The QCD parameters for this fit are
$\rho_A=2$ and $\phi_A=2.75$, where $\rho_A$ and $\phi_A$ are
defined by $X_{A} \equiv \int^1_0 {dx \over x}  \equiv \left( 1 +
\rho_{A} e^{i \phi_{A}} \right)
 {\rm ln} {m_B \over \Lambda_h}$~\cite{bbns}.
The ratio $\Delta M_{B_s}/\Delta M_{B_d}$ is 34.3 for this model
point.  The EDMs are following: $|d_e| =2.48 \times 10^{-29}$~e~cm
and $|d_n| =8.6 \times 10^{-28}$~e~cm.  The BR of $b \to s\gamma$ is
$4.2 \times 10^{-4}$.

\begin{table}
\caption{The CP-averaged branching ratios ($\bar {\cal B}$ in units
of $10^{-6}$), the direct CP asymmetries (${\cal A}_{CP}$), and the
effective $\sin (2\beta)$ for $m_{1/2}=450$ GeV, $m_0=300$ GeV and
$\tan\beta=40$, $|{\Delta_{23}}_{LL}|=0.48$, $|\Delta
A^d_{23}|=0.1$, $|\Delta A^u_{23}|=0.3$.}
\smallskip
\begin{tabular}{|c|c||c|c|}
\hline
 BR & Average & CP asymmetry & Average  \\
\hline $\bar {\cal B}(B^{\pm} \to K^0 \pi^{\pm})$ & $23.8$
 & ${\cal A}_{CP}(K^0 \pi^{\pm})$ & $-0.03$  \\
$\bar {\cal B}(B^{\pm} \to K^{\pm} \pi^0)$ & $11.1$
 & ${\cal A}_{CP}(K^{\pm} \pi^0)$ & $0.013$  \\
$\bar {\cal B}(B^0 \to K^{\pm} \pi^{\mp})$ & $19.6$
 & ${\cal A}_{CP}(K^{\pm} \pi^{\mp})$ & $-0.10$  \\
$\bar {\cal B}(B^0 \to K^0 \pi^0)$ & $11.4$
 & ${\cal A}_{CP}(K^0 \pi^0)$ & $-0.11$  \\
 &  & $\sin (2\beta^{\rm eff})_{K_s \pi^0}$ & $+0.8$  \\
\hline $\bar {\cal B}(B^{\pm} \to \eta' K^{\pm})$ & $72$
 & $\sin (2\beta^{\rm eff})_{\eta' K^0}$ & $+0.6$  \\
\hline
\end{tabular}
\label{table:2}
\end{table}

The origin of the $(m^2_{LL,RR})_{23}$ terms are natural in the
grand unifying models which explain neutrino masses.  For example,
if right handed neutrinos exist, SU(5) might generate the term
$Y_{\nu}\bar 5 {\bar N} 5_H$, where $\bar {5}$ has $d^c_i$ and the
lepton $L$ doublet, ${\bar N}$ is the singlet right handed neutrinos
and $5_H$ contains the SM Higgs doublet (along with the colored
Higgs fields. $Y_{\nu}$ has a flavor structure in order to explain
the neutrino masses and bilarge-mixing angles.  Now these couplings
introduce flavor violation to the soft masses ($\tilde d^c$ and
$\tilde l$) via the RGEs, ${d m^2\over{dt}}\propto m^2 Y_{\nu}
Y_{\nu}^{\dag}$.  In this model the $m_{ij,\bar{5}}^2$ terms for
$i\neq j$ can be generated~\cite{Hisano:2003bd}. These terms get
introduced between the GUT scale and the string scale due to the
RGEs. One expects these flavor violating terms also in the SO(10)
type models~\cite{Dutta:2005ni}. The right handed neutrinos there
belong to the fundamental 16 representation of SO(10) and produce
these flavor violating terms in the soft masses.  The flavor
structures of the Dirac and Majorana coupling arise from the
neutrino mixing matrix.  The $A_{ij}$ terms (for $i\neq j$) also get
contributions from the flavor structure of  $Y_{\nu}$ due to the
quark-lepton unification.  Similar flavor violating effects in the
soft terms are also present in the Pati-Salam type
models~\cite{Deshpande:1996jv}. In this case, the quark-lepton
unification can happen at the intermediate scale and the flavor
violating Majorana coupling $f \psi_{{\bar 4},1,2}\psi_{{\bar
4},1,2} \Delta_{10,1,3}$ ($\psi_{{\bar 4},1,2}$ contains right
handed neutrinos along with the right handed quarks and leptons,
$\Delta_{10,1,3}$ is the new Higgs field ) is responsible for right
handed neutrino Majorana masses.  Now the RGEs involving these
couplings between the intermediate scale and the grand unifying
scale can easily introduce  flavor violating terms in the squark and
the slepton masses.

In conclusion, we find that the current experimental results on the neutral $B_s$ meson mass difference
have introduced strict constraint on the the SUGRA parameter space for flavor mixing terms $A_{23}$ and
$(m^2_{LL,RR})_{23}$ in the soft SUSY breaking terms.  These flavor violating soft breaking terms are
natural in the grand unifying models.  In order to explain the $B\to K\pi$ puzzle, $A_{23}$ and
$(m^2_{LL,RR})_{23}$ are needed.  We show that it is still possible to explain the $B\to K\pi$ puzzle even
after satisfying the new Tevatron result on $\Delta M_{B_s}$.
The model used here contains three complex nonuniversal soft breaking terms ($\Delta_{23,LL}$,
$\Delta A^{u,d}_{23}$), though an acceptable fit can be obtained using just $\Delta A^{u,d}_{23}$.
This allows us to calculate 19 observables of the $B$ system (9 observables in the $B\rightarrow\pi K$
modes, 4  observables in $B\rightarrow\phi K$ modes, 5 observables in the $B\rightarrow\eta^{(\prime)} K$
modes and ${\cal B}(b\rightarrow s\gamma)$).   The future results on $\sin (2\beta^{\rm eff})_{K_s \pi^0}$
and $A_{CP}(K^{\pm} \pi^{\mp})$ are crucial to probe this model.  Finally, we find that the large
${\cal B}(B\to\eta'K)$ can be explained in this parameter space with  $\sin (2\beta^{\rm eff})_{\eta' K^0}$
near the current experimental result which is 2$\sigma$ away from $\sin (2\beta^{\rm eff})_{J/\psi K}$~.

\vspace{1cm}
\centerline{\bf ACKNOWLEDGEMENTS}
\noindent
The work of B.H. was supported in part by the National Nature Science Foundation of China
(No. 10505011). The work of S.O. was supported by the Korea Research Foundation Grant
(KRF-2004-050-C00005).


\end{document}